\documentclass[conference]{IEEEtran}
\usepackage[utf8]{inputenc}
\IEEEoverridecommandlockouts
\usepackage{cite}
\usepackage{amsmath,amssymb,amsfonts}
\usepackage{graphicx}
\usepackage{textcomp}
\usepackage{xcolor}
\usepackage{algorithm}
\usepackage{algorithmicx}
\usepackage{algpseudocode}
\usepackage{multirow}
\usepackage{float}
\usepackage{caption}
\usepackage{multirow}
\usepackage{diagbox}
\usepackage{booktabs}
\usepackage{subfigure}
\def\BibTeX{{\rm B\kern-.05em{\sc i\kern-.025em b}\kern-.08em
		T\kern-.1667em\lower.7ex\hbox{E}\kern-.125emX}}
\begin{document}
	
\title{A Matrix Decomposition Model Based on Feature Factors in Movie Recommendation System\\
	}
	
	\author{\IEEEauthorblockN{1\textsuperscript{st} Dan Liu}
		\IEEEauthorblockA{\textit{School of mathematical science} \\
			\textit{University of Electronic Science and Technology of China}\\
			email: 2867135435@qq.com}
		\and
		\IEEEauthorblockN{2\textsuperscript{nd} Hou-biao Li}
		\IEEEauthorblockA{\textit{School of mathematical science} \\
			\textit{University of Electronic Science and Technology of China}\\
			email: lihoubiao0189@163.com}
	}
	
	\maketitle
	
	\begin{abstract}

Currently, matrix decomposition is one of the most widely used collaborative filtering algorithms by using factor decomposition to effectively deal with large-scale rating matrix. It mainly uses the interaction records between users and items to predict ratings. Based on the characteristic attributes of items and users, this paper proposes a new UISVD++ model that fuses the type attributes of movies and the age attributes of users into SVD++ framework. By projecting the age attribute into the user's implicit space and the type attribute into the item's implicit space, the model enriches the side information of the users and items. At last, we conduct comparative experiments on two public data sets, Movielens-100K and Movielens-1M. Experiment results express that the prediction accuracy of this model is better than other baselines in the task of predicting scores. In addition, these results also show that UISVD++ can effectively alleviate the cold start situation.
	\end{abstract}
	\begin{IEEEkeywords}
		Movies recommendation, collaborative filtering, matrix factorization, feature factor cold start, data sparsity
	\end{IEEEkeywords}
	\section{Introduction}
	At present, users are more and more involved in the production of data, which makes data and information explode at an exponential speed, leading to the problem of information overload. The current research in information processing is mainly reflected in two aspects: information filtering and personalization. Massive information and personalized demand promote the emergence and development of recommendation system.
	
	Collaborative filtering (CF) is based on the historical interaction data between users and items, learning users' preference patterns for items, and then automatically identifying the items that users may be interested in from a large number of items that users have not yet interacted with, forming personalized project recommendations. Therefore, collaborative filtering utilizes the "wisdom of crowds"\cite{b1} to generate the user's recommendation list. This recommendation strategy is intuitive and easy to implement, and is suitable for various fields, such as books and movies. In recent years, there have been many algorithmic debates, which have been widely used \cite{b2,b3,b4,b5}. However, dealing with the data sparsity and cold start of the scoring matrix remains a thorny challenge for recommendation systems.  When a new user/item joins the system, there is only a small amount of interactive information, which can cause cold start problems\cite{b6}.
	
	In order to effectively solve similar problems, researchers focused on the extraction methods of auxiliary information such as features, then combined with the traditional recommendation algorithm, and proposed a variety of hybrid CF methods. Among all kinds of algorithms, matrix factorization (MF) is the most successful CF algorithm due to its flexibility and expansibility\cite{b7}. The traditional matrix decomposition model is considered to use edge information to improve the recommendation accuracy\cite{b8},\cite{b9}. However, due to the limited learning ability and sparse side information, these methods are difficult to grasp the deep relationship between users and items. Therefore,  user-side auxiliary information (comment feedback, etc.) and item-side auxiliary information (project characteristics, etc.) are widely used for mining and application to improve recommendation performance \cite{b10,b12}. Recently, auxiliary information has been widely applied in the recommendation system, and great achievements have been achieved in the rating prediction task.

	Based on these methods, this paper proposes a new model -- UISVD++, which learns related attributes from the auxiliary information of users and items, projects the user's age and movie type attributes into the same potential space as the user and item, and then integrates into the SVD++ framework. Finally, we use two real-world datasets to conduct performance comparison experiments\cite{b13}. The main contributions of our study are summarized as follows:

1) We analyze the existing methods to solve data-sparsity and cold-start, and propose a collaborative filtering model--UISVD++. This model can not only take the implicit relationship between users and items, but also represent the extracted attributes of users and items as vectors, which share the common potential space with users and items respectively.

2) We conduct comprehensive experiments with UISVD++ model on datasets, and the experimental results show that the UISVD++ model performed better than the experimental baseline in the scoring prediction task. At the same time, the model is also proved to be effective in alleviating data sparseness and cold start situation.

The rest of this article is set as follows. Section II systematically describes MF technology and its related improvements, incorporating additional information into MF framework and deep learning methods. Section III covers the basics to be used in this paper.  Section IV describes in detail our proposed UISVD++ model. Section V gives the comparison test and experimental results, and carries on the analysis to them. Finally, section VI contains the summary of this paper and the future research direction.

\section{Related Work}
As everyone knows, CF makes recommendations to other users based on the utilization of the interaction between users and items, and effectively identifies the preferred items of specific users. However, the performance of CF is often challenged in the cases of sparse-data and cold-start. In these cases, many researchers have focused on improving recommendation performance by utilizing auxiliary information from users and items \cite{b4}, \cite{b5}. For example, some recommendation systems used social networks \cite{b14}, users' comments \cite{b10}, \cite{b11} to extract relevant auxiliary information and integrate it into MF framework to improve recommendation performance. In recent years, with the wide application and rapid development of deep learning technology, based on its advantages in feature extraction, deep learning has been used to learn potential representations in recommendation systems\cite{b16}. However, recent research results show that the calculations of models based on deep learning are very complex and fall short of the results expected by researchers \cite{b17}, \cite{b18}. In contrast, matrix factorization (MF) remains one of the most popular CF approaches because it allows for the integration of additional information.

The MF method uses the decomposition of user-item rating matrix to express user interest and item features as potential factor vectors in the common potential space. After calculating the inner product, the task of predicting scores and recommending is completed\cite{b19}. Simon Funk only considers the existing scoring records and puts forward a model (FunkSVD) in \cite{b20}, which can learn user and item latent matrix $P$ and $Q$ from user-item rating information and predict user's rating of commodities by reconstructing low dimensional matrix. With the Netflix prize competition, the SVD concept promotes the emergence of other recommendation models, such as SVD++\cite{b21}, PMF\cite{b22}, which are still often used as reference baselines for MF-based recommendation studies and are also the core element of using user/item related information to form a hybrid model\cite{b14}.

The matrix factorization collaborative filtering recommendation based on review proposed by Duan, based on users' comments on the items, a rating matrix is established after quantifying the sentiment and interpolated into the matrix factorization method to alleviate the problem of data sparsity\cite{b23}. Hai Liu proposed EDMF model, which extracted features in user comments by convolutional neural network and attention mechanism technology, constrained the sparsity of comment text, and integrated MF technology to improve the effectiveness and efficiency of the model\cite{b24}. Lai\cite{b25} used the user's knowledge graph to construct the user's preference vector, integrated it into the matrix factorization algorithm, and proposed a new recommendation model, which effectively improved the accuracy of the recommendation system. Zhang proposed the FeatureMF model, which projected the available attributes of items into the same latent space as users and items, and integrated into the matrix factorization framework to alleviate the cold-start problem of recommendation systems\cite{b26}. Wang and Hong\cite{b27} proposed the MFFR model, which used LDA topic modeling to obtain the user's comment partial partition matrix, and then integrated it into the MF model to alleviate the problem of data sparsity. Zhu and Yan proposed the TFRMF model, which used deep learning and other means to obtain features from the comment text, and then introduced them into the PMF model to alleviate the cold start problem\cite{b28}.

Autoencoder model is also commonly used to solve the cold-start problem of recommender systems. Suvash Sedhain applied autoencoder in collaborative filtering recommendation and proposed a new AutoRec framework, which combined with users and items respectively to generate U-Autorec and I-Autorec models\cite{b29}. Zheng et al. combined user-based and item-based autoencoders into a framework, combined the advantages of the two encoders, and proposed a new collaborative filtering method based on Autoencoders (UIAE)\cite{b30}. To alleviate the problem of data sparsity in the recommendation system, Zhang et al. proposed SSAERec algorithm, which extracted the features of items by stacking sparse encoders and integrated the features into the matrix factorization framework\cite{b31}. Darban et al. constructed a similarity map to extract the user's feature information, and then combined it with the user's basic features (age and gender). After extracting the features and reducing the dimension by autoencoder, they recommended the user by clustering method, and named the model GHRS\cite{b32}.
Through the analysis of the existing recommendation technology, we find that using side information can effectively improve the accuracy of the recommendation model. Hence, this paper proposes a new recommendation model, which integrates user and item characteristic attributes into SVD++ framework. During the experiment, we extracted the user's age and item's type attributes from the datasets, and represented each attribute as a potential vector, which was projected into the same potential factor space as the user and item.
\section{Relevant Knowledge}
In this section, we will introduce the relevant characteristic attributes and the basic knowledge of matrix decomposition to be used in this paper. We will start with the notations used in the rest of this article, and then briefly introduce MF.

\subsection{Notations and Nomenclature}\label{AA}
As for the potential information of the user, some characteristics of the user can be obtained from the auxiliary information on the user side. Each feature of a user may contain different attributes. For example, user information contains a variety of characteristics such as gender and age, while gender can include both male and female attributes. In this paper, it is assumed that the attribute information of user $u$ can be used as a sequence $F = \left\{ {{F_{1_j}},{F_{2_j}}, \ldots ,{F_{{M_j}}}} \right\}$, where $M_j$ is the total number of  attributes corresponding to user feature $j$. In this paper, we use MovieLens-100K and MovieLens-1M datasets to collect the age feature $F_a$. For feature $F_a$, the attribute set belonging to the user $u \in U$ is labeled as ${F_a}\left( u\right)$. Table I shows the attributes of user in the MovieLens dataset.
\begin{table}[h]
	\vspace{10pt}
	\caption{Characters of the users.}
	\centering
	\begin{tabular}{ | l | l | l | }
		\hline
		\multicolumn{3}{|c|}{Users} \\
		\cline{1-3}
		Gender & Occupation & Age \\
		\hline
		Female  & Educator & 22 \\
		Male   & Writer  & 34 \\
		Female   & Actor &  54 \\
		\hline
	\end{tabular}
\end{table}

Similar to the user potential information, the item side can also extract the relevant feature attributes. For instance, the Movielens dataset collects actors and types characteristics of films, while item features may contain attributes such as Adventure and Comedy. In this paper, the underlying information of user $i\left( {i \in I} \right)$ can also be used as an array $F = \left\{ {{F_{1_i}},{F_{2_i}}, \ldots ,{F_{{N_i}}}} \right\}$, where $N_i$ is the total number of attributes corresponding to item feature $i$. To make better use of these features, this article adds a set of attributes for the film type feature $F_t$. Table II shows the attributes of films in the MovieLens dataset. The other notations in this article are shown in Table III.
\begin{table}[h]
	\vspace{10pt}
	\caption{Characters of the movies.}
	\centering
	\begin{tabular}{ | l | l | l | }
		\hline
		\multicolumn{3}{|c|}{Items} \\
		\cline{1-3}
		Directors & Actors & Types  \\
		\hline
		Bille August  & Sandra Staggs & Adventure \\
		Chris Wedge   & Jelena Nik  & Comedy \\
		Coppola   & Roy Abramsohn &  Romance \\
		\hline
	\end{tabular}
\end{table}

\begin{table}[h]
	\vspace{10pt}
	\caption{Nomenclature.}
	\centering
	\begin{tabular}{ l | l }
		\hline
		Symbols & Definitions \\
		\hline
		$U$  & The user set, $ U = \{ {u_1},{u_2}, \ldots ,{u_m}\}$. \\
		$I$   & The item set, $ I = \{ {i_1},{i_2}, \ldots ,{i_n}\}$.  \\
		$K$ & The set of exposure ratings.\\
		${r_{ui}}$   & User $u$'s ratings of item $i$.   \\
		${\tilde r_{ui}}$   & User $u$'s predicted rating of the item $i$.  \\
		$m$   & Number of $U$.  \\
		$n$   & Number of $I$.  \\
		$k$   & Dimension of potential feature space.  \\
		${p_u}/{q_i}$   & The latent characteristics of user $u$ / item $i$.  \\
		${b_u}/{b_i}$   & The deviation coefficient of user $i$ / item $u$.   \\
		$\mu$   & Global average ratings.  \\
		$\lambda$ & Regularization parameter. \\
		$F_a$   & Set of feature for the users' age.  \\
		$F_t$   & Set of feature for the movies' type.  \\
		$\left| {{F_a}} \right|$   & Number of $F_a$.   \\
		$\left| {{F_t}} \right|$   & Number of $F_t$.  \\
		${y_a}$   & Vector representing of  $F_a$ and ${y_a} \in {\mathbb{R}_{\left| {{F_a}} \right| \times d}}$. \\
\		${y_t}$   & Vector representing of  $F_t$ and ${y_t} \in {\mathbb{R}_{\left| {{F_t}} \right| \times d}}$. \\
		\hline
	\end{tabular}
\end{table}
\subsection{One-Hot}
Hot vector encoding is a simple way to create sparse vectors from text. Each element of the vector corresponds to a word in the entire corpus. A heat vector contains mostly zeros, with the exception of elements corresponding to words occurring in the text, which are set to 1. These vectors are simple positional encodings that do not contain any semantic content about the word except that it is equivalent or not equivalent to other words in the corpus.

One-Hot coding (One-Hot) is to encode the N states of the field with N bit state registers, so that each state has an independent register bit. For example, if the "category" of a program in the data set has only three values, it can be encoded with three binary bits, namely 100, 010, 001. When each field is valued, only one binary bit is activated, and the three codes correspond to culture, entertainment, and sports respectively.

\subsection{Matrix Factorization}
Recommendation system usually consists of three parts: a set of users $U = \{ {u_1},{u_2}, \ldots ,{u_m}\}$, a set of items $I = \{ {i_1},{i_2}, \ldots ,{i_n}\}$ and a ratings matrix $R\left(R \in {\mathbb{R}_{m \times n}} \right)$ composed of them. The matrix decomposition model is to decompose the ratings matrix $R$ into the product of $m \times k$ dimensional users’ implicit factor space $P$ and $n \times k$ dimensional items’ implicit factor space $Q$, and make the inner product of implicit vector $p_u$ and $q_i$ as close as possible to the original score $r_{ui}$.

SVD++ matrix decomposition model takes into account both the fixed attributes of users and items and the implicit feedback of users, so the implicit feedback and user attributes are introduced while the deviation between users and items is introduced. That is, add user preferences on the $p_u$. Suppose that the sets of objects in the scoring matrix also have a hidden factor vector, which is the same dimension of the hidden factor vector in the object. Add up the hidden factor vector of items that the user has scored to express the user's preferences. The prediction ratings function is
\begin{equation}
	{\tilde r_{ui}} = \mu  + {b_u} + {b_i} + ( {{p_u} + {{\left| {N\left( u \right)} \right|}^{ - \frac{1}{2}}}\sum\limits_{j \in N\left( u \right)} {{y_j}} } )q_i^T.
\end{equation}
where $\mu$ is the global average rating, $b_u$ and $b_i$ indicate the deviation coefficient. The initial value of $b_u$ and $b_i$ can be set to a zero vector in the optimization.

To improve the prediction accuracy, a regular term is introduced in the loss function to avoid the over-fitting problem. The loss function of SVD++ is as follows
\begin{align}
	{O_{SVD++}} = \sum\limits_{\left( {u,i} \right) \in K} {{{\left( {{r_{ui}} - {{\tilde r}_{ui}}} \right)}^2}}+\lambda ({{{\left\| {{p_u}} \right\|}^2} + {{\left\| {{q_i}} \right\|}^2}}
	\notag
	\\
	\phantom{=\;\;}
	+ {b_u}^2 + {b_i}^2 + {\sum\limits_{j \in {N_u}} {\left\| {{y_j}} \right\|} ^2})
\end{align}
where ${\left\| \cdot  \right\|}$ is the Frobenius norm \cite{b31}.

\section{The Proposed Model}
In this section, we will describe our proposed matrix decomposition model based on the fusion of project and user characteristics.

\subsection{General Model Overview}\label{AA}
UI\_SVD$++$ is the assumption that when predicting the score of the user $u$ and the item $i$, it depends not only on $u$ and $i$, but also on the feature properties of the user $u$ and the feature attributes of the item $i$. Different from previous studies, we integrate these side information into the SVD++ framework in the form of vectors. In order to conform to the hypothesis of this paper, we propose to encode the available attributes of users and items into multi-dimensional vectors by One-hot coding technology, and map these side information into the same latent factor space as users and items respectively, and then integrate them into the matrix factorization framework. In general, the UISVD++ model contains both user and item aspects. On the user side, UISVD++ is composed of two parts: 1. User-specific latent factors and 2. Attributes available on the user side. In terms of item, UISVD++ is composed of two parts: 1. Potential factors specific to the item; 2. Properties available on the item side. Figure 1 is the structure diagram of UISVD++ overall mode.

\begin{figure*}[!t]
	\begin{center}
		{
			\includegraphics[width=7in,height=3.5in]{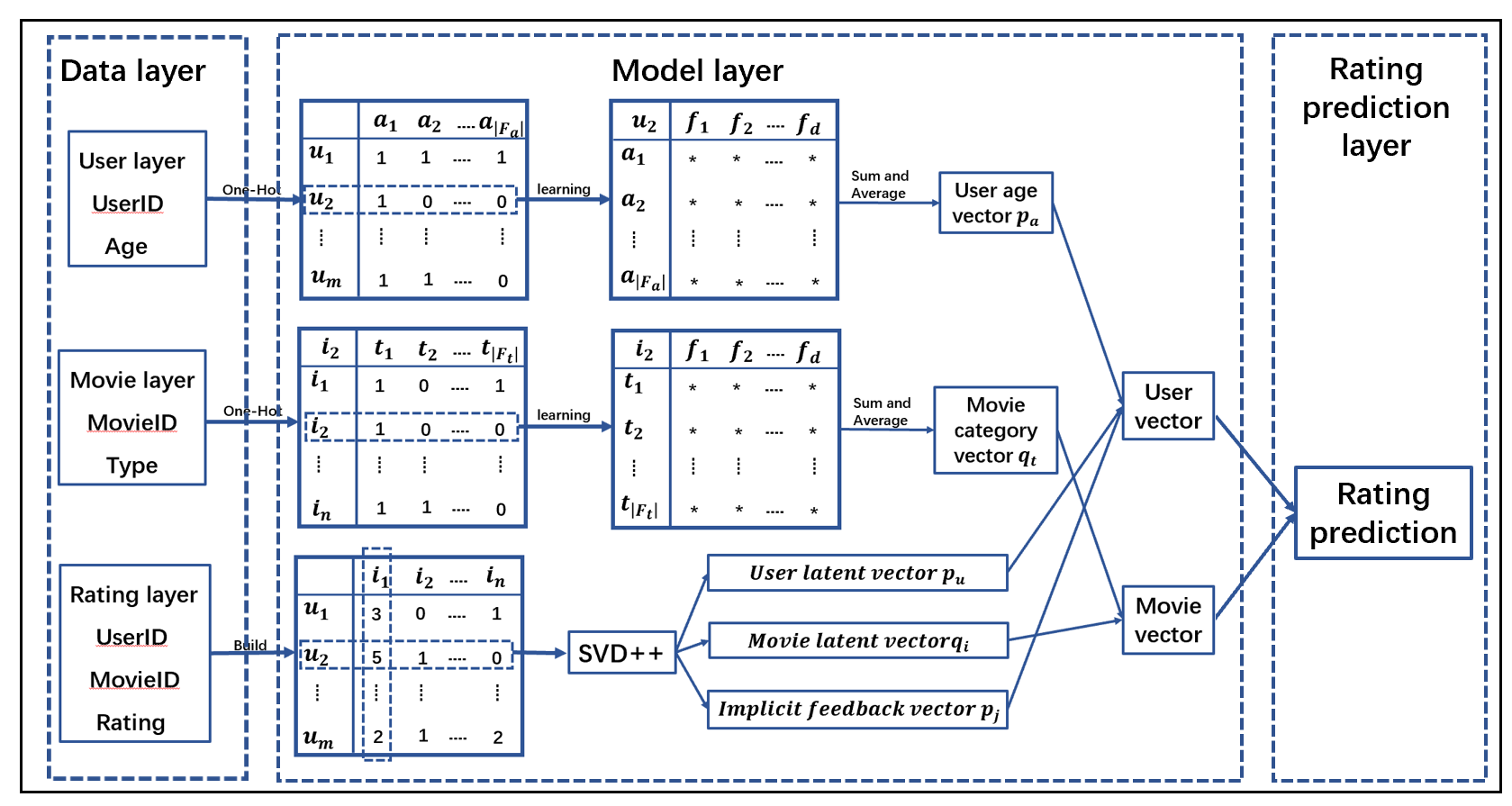}
		}
		\caption{Schematic diagram of the structure of the UISVD++ overall model.}
	\end{center}
\end{figure*}	
\subsection{Side Information Processing}\label{AA}
In this paper, let ${y_a}\left\{ {{y_a}\left| {{y_a} \in {\mathbb{R}_{\left| {{F_a}} \right| \times d}}} \right.} \right\}$ be the set of user's age features which used for movie recommendations, where ${y_a}$ represents a matrix representation for the age attribute of the user $u$, $\left| {{F_a}} \right|$is the number of attributes that represent a type feature. In the ML-1M dataset, there have been studies that have divided the age of users into 7 attributions according to their age, which are 0-18 years old, 18-24 years old, 25-34 years old, 35-44 years old, 45-49 years old, 50-55 years old and above 56 years old. Therefore, this paper adopts the same decomposition method to decompose the age in the data set to the same attribution. Then, using One-hot coding technique, each age stage is projected as 0 and 1 respectively. If it is in this age stage, we encode this stage as 1, otherwise it is 0. Based on these attributes, each user's age has a corresponding potential factor vector. In subsequent calculations, we project this vector into the user's latent factor space. Table IV illustrates vectors of the user's age. The user's potential vectors are averaged and then integrated into the MF framework.
\begin{table}[h]
	\vspace{10pt}
	\caption{Latent factor vectors of users for UISVD++.}
	\centering
	\begin{tabular}{ l | l | l}
		\hline
		user & age &vector \\
		\hline
		u1  & 17 & 1100000 \\
		u2  & 24 & 1110000  \\
		u3  & 9 & 1000000   \\
		u4  & 56 & 1111110  \\
		...  & ... & ...  \\
		\hline
	\end{tabular}
\end{table}

Let ${y_t}\left\{ {{y_t}\left| {{y_t} \in {\mathbb{R}_{\left| {{F_t}} \right| \times d}}} \right.} \right\}$ be the set of movie genre features in a movie recommendation, where ${y_t}$ is the matrix representing for the type attribute of item $i$, $F_t$ is the number of attributes that represent a type feature. The One-hot encoding method is also used to encode the project side information (type, actor, etc.) into vectors. Therefore, the UI\_ SVD$++$ consists of two parts for item $u$: 1. The user-specific latent factors $p_u$. 2. Sum of the attribute of user's age in each of the features that user $u$ possesses${\left| {{F_a}} \right|^{ - 1}}\sum\nolimits_{i \in {F_a}} {{y_a}(i)} $.

Similarly, UI\_SVD++ for item $i$ is also composed of two parts.
Hence, the prediction method of UISVD++ model can be described by mathematical symbols as
\begin{equation}
	{\tilde r_{ui}} = \mu  + {b_u} + {b_i} + {\left( {{p_u} + {p_a}} \right)}\left( {{q_i} + {q_t}} \right)^T,
\end{equation}	
where ${p_a}$ represents the age profile of the user ${p_a} = {\left| {{F_a}} \right|^{ - 1}}\sum\nolimits_{i \in {F_a}} {{y_a}(i)}$. ${q_t} = {\left| {{F_t}} \right|^{ - 1}}\sum\nolimits_{i \in {F_t}} {{y_t}(i)}$ is the type features of film. The definitions of $\mu$, ${b_u}$ and ${b_i}$ are shown in Table III.

Apparently, UISVD++ is able to predict ratings based solely on the age of the user even in the case of a cold start, alleviating the situation of a cold start. Similarly, item types can be used to predict relevant ratings and alleviate cold start situations of items.

Although the model introduced feature representation and extracted auxiliary information of users and items, the performance of the algorithm was still not ideal and could not produce the expected recommendation accuracy. For that reason, we combine this model with SVD++ model to become a complete UISVD++ model. This algorithm can also take into account the implicit feedback information of users and items to alleviate the problem of over-sparse data. Therefore, UISVD++ integrates the user age and film type feature vectors to predict the user rating formula as follows
\begin{equation}
	{\tilde r_{ui}} = \mu  + {b_u} + {b_i} + {\left( {{p_u} + \alpha {p_a} + \beta {p_j}} \right)}\left( {{q_i} + {q_t}} \right)^T,
\end{equation}	
where ${p_j} = {\left| {N\left( u \right)} \right|^{ - \frac{1}{2}}}\sum\nolimits_{j \in N\left( u \right)} {{y_j}}$ is the implicit behavior of a user towards an item.$ \alpha$ and $\beta$ are the weight coefficient of user implicit feedback and age attribute, and $\alpha+\beta = 1$.

\subsection{Optimization of Model}
Our model trains model parameters by minimizing regularization squared error loss, and the loss function of UISVD++ is
\begin{equation}
Loss = \sum\limits_{\left( {u,i} \right) \in K} {{{\left( {{r_{ui}} - {{\tilde r}_{ui}}} \right)}^2} + } \lambda  \cdot Z,
\end{equation}
where $ \lambda$ is the learning rate, and $Z$ represents the regularization term used to prevent over-fitting and gradient explosion problems. $Z$ is expressed as
\begin{equation}
	\begin{split}
Z = &{\left\| {{p_u}} \right\|^2} + {\left\| {{q_i}} \right\|^2} + {b_u}^2 + {b_i}^2 + \sum\limits_{i \in {F_a}} {{{\left\| {{y_a}\left( i \right)} \right\|}^2}}  \\
 &+ {\sum\limits_{j \in {N_u}} {\left\| {{y_j}} \right\|} ^2} +\sum\limits_{i \in {F_t}} {{{\left\| {{y_t}\left( i \right)} \right\|}^2}} \,
\end{split}
\end{equation}
where, ${\left\|  \cdot  \right\|}$ is the Frobenius norm \cite{b31}.

For the optimization of the model, stochastic gradient descent (SGD) is used to search for local minima because the optimization problem in the above equation is doubly convex. For a given training dataset instance, the parameters in the opposite direction of the gradient are updated, and an iterative cycle is performed over all observation ratings. Learning the pseudo-code of UISVD++ model is shown in algorithm 1.
\begin{algorithm}[htb]
	\caption{UISVD++ Model}
	\label{alg:Framwork}
	{\bf Input:} R, $\lambda $, $\alpha $, $\beta $, $\gamma $, $k$, $step \leftarrow 0$.\\
	{\bf Output:} predicted rating: ${\tilde r_{ui}}$.
	\begin{algorithmic}[1]
		\State Initialize the bias vectors  $b_u$ and $b_i$, implicit factor matrix $p$ and $q$, weight matrix $y_j$ for implicit feedback calculation,  user age matrix $y_a$ and film type matrix $y_t$;
		\While {$step<n\_step$ }
		\While{$(u,i) \in K$}
		\State  ${\tilde r_{ui}} \leftarrow \mu  + {b_u} + {b_i} + {\left( {{p_u} + {p_a} + {p_j}} \right)^T}\left( {{q_i} + {q_t}} \right)$	
		\State ${e_{ui}} \leftarrow {r_{ui}} - {\tilde r_{ui}}$
		\State ${b_u} \leftarrow {b_u} - \gamma \left( {{e_{ui}} + \lambda {b_u}} \right)$
		\State ${b_i} \leftarrow {b_i} - \gamma \left( {{e_{ui}} + \lambda {b_i}} \right)$
		\State${p_u} \leftarrow {p_u} - \gamma \left( {{e_{ui}}\left( {{q_i} + {q_t}} \right) + \lambda {p_u}} \right)$
		\State ${q_i} \leftarrow {q_i} - \gamma \left( {{e_{ui}}\left( {{p_u} + {p_a} + {p_j}} \right) + \lambda {q_i}} \right)$
		\For{$i \in {F_a}$}
		\State$q = {q_i} + {q_t}$
		\State ${y_a}\left( i \right) \leftarrow {y_a} - \gamma \left( {{e_{ui}}\alpha {{\left| {{F_a}} \right|}^{ - 1}}q + \lambda {y_a}\left( i \right)} \right)$
		\EndFor
		\For {$j \in N\left( u \right)$}
		\State ${y_j} \leftarrow {y_j} - \gamma \left( {{e_{ui}}\beta {{\left| {N\left( u \right)} \right|}^{ - \frac{1}{2}}}\left( {{q_i} + {q_t}} \right) + \lambda {y_j}} \right)$
		\EndFor
		\For{$i \in {F_t}$}
		\State $p = {p_u} + \alpha {p_a} + \beta {p_j}$
		\State ${y_t}\left( i \right) \leftarrow {y_t} - \gamma \left( {{e_{ui}}{{\left| {{F_t}} \right|}^{ - 1}}p + \lambda {y_t}\left( i \right)} \right)$
		\EndFor
		\EndWhile
		\State $step \leftarrow step + 1$
		\EndWhile
	\end{algorithmic}
\end{algorithm}
\section{The experiment}

In this section, we will use two benchmark datasets for performance testing of comparative experiments.

\subsection{Experimental establishment}\label{subsecA}
\subsubsection{Datasets}

\indent The datasets we use are MovieLens-100K and MoiveLens-1M \cite{b13}. MovieLens-100K\footnote {https://grouplens.org/datasets/movielens/100k/} dataset contains 1682 films, 943 users and 100,000 rating. Meanwhile, MoiveLens-1M\footnote {https://grouplens.org/datasets/movielens/1m/} dataset covers 3706 films, 6040 users and 1,000,000 rating. Each rating is an integer between 1(be regarded as the worst) and 5(be regarded as the best), and the datasets are very sparse. The statistics of the dataset are shown in Table V.
   \begin{table}[H]
    \vspace{10pt}
   	\caption{Datasets statistics.}
   	\centering
   	\begin{tabular}{ccccc}
   		\toprule
   		Datasets & users & items& ratings & sparsity \\
   		\midrule
   		MovieLens-100K&943&1682&60089&93.7\% \\
   		MoiveLens-1M&6040&3706&1000209&95.5\% \\
   		\bottomrule
   	\end{tabular}
   \end{table}

\subsubsection{Evaluation Measures}

\indent In evaluating the recommendation accuracy of the recommendation system, we use two error-based metrics, namely MAE and RMSE \cite{b32},\cite{b33}, which are defined in (9) and (10) respectively
\begin{equation}
	MAE = \frac{1}{{\left| T \right|}}\sum\limits_{\left( {u,i} \right) \in K} {{{\left| {{r_{ui}} - {{\tilde r}_{ui}}} \right|}}},
\end{equation}
\begin{equation}
RMSE = \sqrt {\frac{1}{{\left| T \right|}}\sum\limits_{\left( {u,i} \right) \in K} {{{\left( {{r_{ui}} - {{\tilde r}_{ui}}} \right)}^2}} },
\end{equation}
where $T$ is the number of fractional exposures in the test set. With the decrease of RMSE and MAE, the accuracy of the model in the prediction task becomes higher.\cite{b34}.
\subsection{Baselines}

In order to test the effect of characteristic factors on SVD++ decomposition, we compare the matrix factorization model of characterization factors with traditional matrix factorization techniques. These models are:
\begin{itemize}	
	\item Bias\_SVD\cite{b5}: a classical matrix decomposition model which considers both user bias and regularization to prevent overfitting.
	\item  FM\cite{b35}: a machine learning algorithm based on matrix decomposition, which takes into account the second order cross features and hidden vectors, and improves to a higher order feature combination.
	\item PMF\cite{b22}: a matrix recommendation algorithm, which uses maximum posterior probability and maximum likelihood estimation for calculation.
	\item SVD$++$\cite{b21}: a matrix decomposition model which contains the advantages of both the neighborhood method and the potential factor method.
	\item FeatureMF\cite{b26}: a matrix decomposition model that projects each attribute data in each item feature into the potential factor space of the users and items.
	\item ConvMF \cite{b28}: a hybrid recommendation model based on deep learning, which firstly extracts features from datasets through convolutional neural network technology, and then integrates them into PMF framework.
	\item DHA-RS\cite{b36}: a deep learning recommendation model combining neural cooperative filtering and stack autoencoder, which can predict user preferences using auxiliary information.
\end{itemize}

Among them, Bias\_SVD, PMF and SVD++ are traditional matrix decomposition models, which are obtained by using the surprise library. The FM, featureMF and ConvMF are matrix decomposition models using features. ConvMF is a context-aware recommendation model combining matrix decomposition and deep learning. DHA-RS is a hybrid recommendation model based on deep learning.

\subsection{Model Performance}
The model presented in this article is run on Python, and part of the experiment uses the Surprise library in Python, a popular Python library that runs recommendation systems. In order to explore the influence of parameters on the UISVD++ model, we used a random subsampling validation method to randomly divide ml-100k and ml-1m datasets\cite{b13} into validation sets and test sets. After several experiments, the average value of test results was displayed.

\subsubsection{Parameter experiment}\label{15}
We randomly divide the datasets into training set (80$\%$) and test set (20$\%$). Use the 50$\%$ discount cross-verification issue. In our experiment, the parameters focus on the user, movie feature dimension k, and regularization coefficients. So, we set the initial value of the learning rate $\lambda$ to 0.01 and the number of iterations to n\_epoch = 20.
{\bfseries Regularization coefficient:} We assume that the number of potential factors is 10 and set the value of the regularization coefficient to be taken from the set $\left\{ {{{10}^{ - 4}},{{10}^{ - 3}}, {{10}^{ - 2}},{{10}^{ - 1}},{1^0},{{10}^1}} \right\}$. One value in, as found by Figure 3, When $\lambda  = {10^{ - 1}}$, the UISVD++ can perform well on datasets ml-100k and ml-1m.

\begin{figure}[H]
	\begin{center}
		\subfigure[ml\_100k]{
			\includegraphics[width=1.5in,height=1.25in]{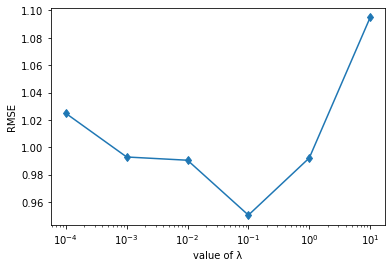}
		}
		\subfigure[ml\_1m]{
			\includegraphics[width=1.5in,height=1.25in]{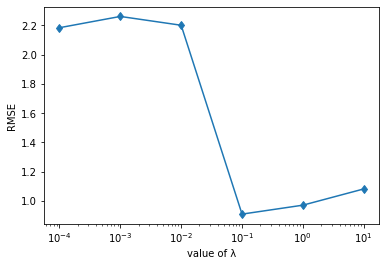}
		}
		\caption{For (a) ml-100k and (b) ml-1m datasets with potential factors of 25, the RMSE performance of UISVD++ varies with the regularization parameter $\lambda$. }
	\end{center}
\end{figure}

{\bfseries Number of latent factors:} As shown in figure 4, different numbers of potential factors have different impacts on the prediction accuracy of the model. When the number of potential factors is 25 and the effect of UISVD++ is the best when applied to the ml-100k datasets. The effect is also the best when UISVD++ is applied on ml-1m datasets and the number of potential factors is 20.
\begin{figure}[H]
	\begin{center}
		\subfigure[ml-100k]{
			\includegraphics[width=1.5in,height=1.25in]{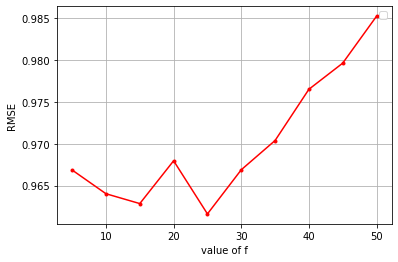}
		}
		\subfigure[ml-1m]{
			\includegraphics[width=1.5in,height=1.25in]{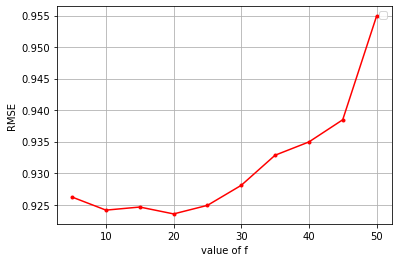}
		}
		\caption{Effect of variation in the number of potential factors on RMSE performance of UISVD++ over two datasets.}
	\end{center}
\end{figure}
{\bfseries Performance convergence:}  When the number of potential factors is set to 25 and the regularization coefficient is set to 0.1, the convergence of ml-100k is shown in Figure 5 (a). It can be found that UISVD++ can achieve better performance when the number of iterations reaches 55. When the number of potential factors is 20 and the regularization coefficient is 0.1, the convergence of ml-1m is shown in Figure 5 (b). It can be found that when the number of iterations reaches the number 50, the best performance can be achieved.
\vspace{-1pt}
\begin{figure}[H]
	\begin{center}
		\subfigure[ml-100k]{
			\includegraphics[width=1.5in,height=1.25in]{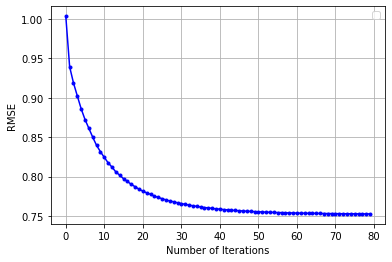}
		}
		\subfigure[ml-1m]{
			\includegraphics[width=1.5in,height=1.25in]{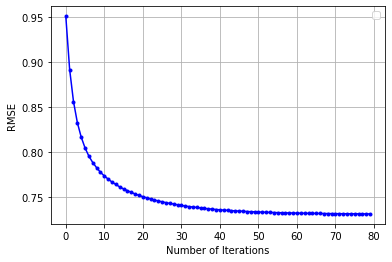}
		}
		\caption{When the number of potential factors is set to 20, the performance convergence of UISVD++ changes with the number of iterations. (a) is ml-100k dataset, (b) is ml-1m dataset. }
	\end{center}
\end{figure}
\subsection{Experimental Results and Analysis}
 In our experiments, we tried to train two datasets with 90\%, 80\% and 50\% data. For data splitting, we randomly split the datasets five times. Finally the average RMSE value of the experimental results is taken as the result. Compared with our UISVD++ model, the experimental results of each model in this paper are shown in Table VI, including the accuracy of UISVD++ on RMSE indexes. The results of the ablation experiment are presented in Table VII, we compare SVD++, USVD++, ISVD++ and UISVD++ on ml-100k and ml-1m dataset with RMSE and MAE metric.
\begin{table*}[!t]
	\caption{Comparison of RMSE performance between UISVD++ and other 7 models in three training scale scenarios on (a) ml-100k and (b) ml-1m datasets.}
	\centering
	\setlength{\tabcolsep}{6mm}{
	\begin{tabular}{|c|c|c|c|c|c|c|}
		\hline
		\multirow{2}{*}{\textbf{Method}}&
		\multicolumn{2}{c|}{\textbf{90\%}}&
		\multicolumn{2}{c|}{\textbf{80\%}}&
		\multicolumn{2}{c|}{\textbf{50\%}}
		\\
		\cline{2-7}
		& \textbf{ml-100k} &\textbf{ml-1m} & \textbf{ml-100k} &\textbf{ml-1m} & \textbf{ml-100k} &\textbf{ml-1m} \\
		\hline
		\multirow{1}*{Bias\_SVD} & 0.9331 & 0.8658 & 0.9435 & 0.8729 & 0.9558 & 0.9039\\
		\multirow{1}*{PMF} & 0.9267 & 0.8934 & 0.9343 & 0.9056 & 0.9498 & 0.9185 \\
		\multirow{1}*{SVD++} & 0.9112 & 0.8556& 0.9219 & 0.8611 & 0.9380 & 0.8861 \\
		\multirow{1}*{FM} & 0.9091 & 0.8561 & 0.9187 & 0.9061 & 0.9350 & 0.8801 \\
		\multirow{1}*{FeatureMF} &0.9194 &0.8576 & 0.9344 & 0.8616&0.9473 & 0.8866 \\
		\multirow{1}*{ConvMF} & 0.9143 & 0.8567 & 0.9167 & 0.8659 & 0.9225 & 0.9012 \\
		\multirow{1}*{DHA-RS} & 0.9132 &0.8521 & 0.9217 & 0.8526 & 0.9270 & 0.8807 \\
		\hline
		\multirow{1}*{\textbf{UISVD++}} &\textbf{0.8989} & \textbf{0.8501}&\textbf{0.9071} &\textbf{0.8514}& \textbf{0.9274} &\textbf{0.8788} \\
		\hline
	\end{tabular}
}
\end{table*}
\begin{figure*}[!t]
	\begin{center}
		\subfigure[ml-100k]{
			\includegraphics[width=3.0in,height=2.25in]{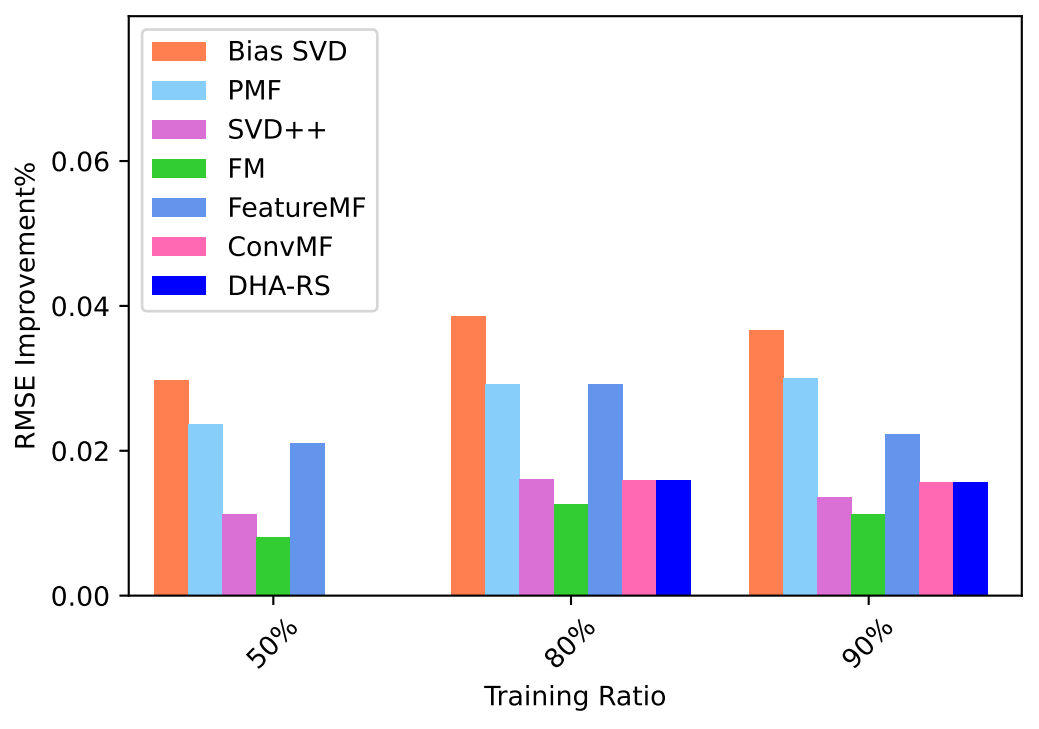}
		}
		\subfigure[ml-1m]{
			\includegraphics[width=3.0in,height=2.25in]{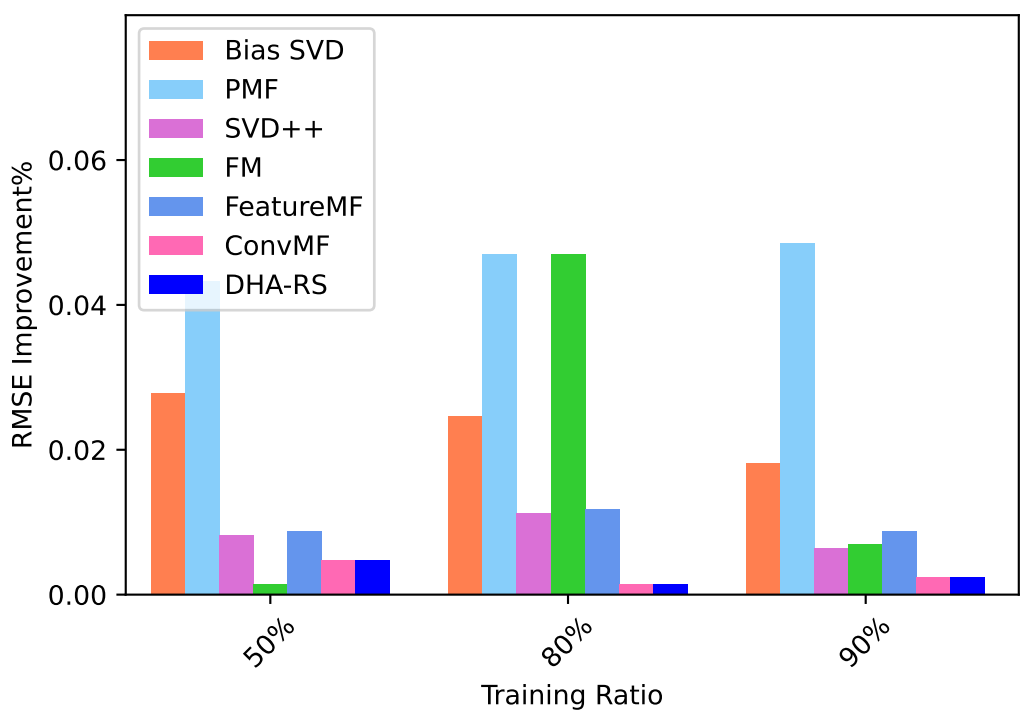}
		}
		\caption{RMSE improvement of UISVD++ performance compared to other models for (a) ml-100k and (b) ml-1m datasets.}
	\end{center}
\end{figure*}
From the above experimental results, we can draw the following conclusions:

\begin{itemize}
	\item [1)]
	Table VI shows the performance of all models on RMSE measures. The results show that our UISVD++ algorithm has lower RMSE on ml-1m and ml-100k datasets, and performs better than the other seven models. When the ratio of the training dataset was 80\%, the RMSR metrics of UISVD++ decreased to 0.9071 and 0.8514 on the above two datasets. And when the training ratio of ml-100k dataset is 80\%, the improvement rate is 1.05\% and 1.58\% compared with ConvMF and DHA-RS. Experimental results show that the method of adding feature factor directly to matrix decomposition can effectively improve the recommendation performance. This validity is derived from the fact that the model can directly learn the characteristics of the goods, the user's attributes, and the implicit feedback of the user, without the need for the intermediate computing or information extraction phases of models such as ConvMF.
	\item [2)]
	Figure 6 shows the RMSE improvement rate of our model compared to other algorithms. The experimental results show that the RMSE values of UISVD++ model on ml-100k and ml-1m are lower than the baseline. The RMSE index of UISVD++ model on ml-1m dataset decreased to 0.8501, 0.8514 and 0.8788, respectively. Compared with the PMF model, UISVD++ improves 4.85\%, 5.99\% and 4.32\% on ml-1m dataset, respectively. However, when the training ratio of ml-100k is 50\%, both ConvMF and DHA-RS are more stable than UISVD++ model. In summary, the experimental results show that as the percentage of data in the training set decreases, the cold start problem becomes more obvious and the data becomes more sparse. The UISVD++ model proposed by us can effectively alleviate data sparsity and cold start problems, and also make more effective use of side information to obtain better recommendation performance.
\end{itemize}
\section{Question}
\subsection{Q1:Function of user side and project side information fusion.}
In order to explore the significance of adding user and item attributes to UISVD++, this paper designed an ablation experiment and compared the UISVD++ model with the following three algorithms:
\begin{itemize}	
	\item SVD++\cite{b21}:a matrix decomposition model which contains the advantages of both the neighborhood method and the potential factor method.
	\item UISVD++: the model combines user features and movie features and integrates them into the SVD++ algorithm.
	\item USVD++: the model only integrates user features into SVD++ algorithm, which is the first deformation of UISVD++.
	\item ISVD++: the model only integrates item features into SVD++ algorithm, which is the second deformation of UISVD++.
\end{itemize}

	Table VIII shows the results of the ablation experiment, where RMSE and MAE were used for comparison. According to analysis, compared with the traditional matrix decomposition method SVD++, the performance of USVD++, ISVD++ and UISVD++ models is improved.  Meanwhile, the RMSE index of SVD++ and UISVD++ decreases from 0.9255 to 0.9057, and the improvement rates of the two datasets are 2.14\% and 1.10\% respectively, indicating the validity of the proposed UISVD++ fusion feature model. It can be found that the matrix decomposition method with auxiliary information of both users and items is superior to the matrix decomposition method which only considers implicit feedback of users, and is superior to the matrix decomposition method which only adopts one auxiliary information.
\begin{table}
	\vspace{8pt}
	\caption{The mean RMSE and MAE values of ablation experiment on ml-100k and ml-1m datasets.}
	\centering
	\begin{tabular}{ccccc}
		\toprule
		\textbf{Method}&Metrics&ml\_100k&ml\_1m\\
		\midrule
		\textbf{SVD++}&RMSE&0.9219&0.8599\\
		&MAE&0.7252&0.6712\\
		\hline
		\textbf{USVD++}&RMSE&0.9105&0.8560 \\
		&MAE&0.7163&0.6699\\
		\hline
		\textbf{ISVD++}&RMSE&0.9213&0.8523\\
		&MAE&0.7262&0.6709\\
		\hline
		\textbf{UISVD++}&RMSE&\textbf{0.9071}&\textbf{0.8514}\\
		&MAE&\textbf{0.7159}&\textbf{0.6692}\\
		\bottomrule
	\end{tabular}
\end{table}
\begin{table*}[htp]
	\vspace{15pt}
	\caption{Each age group corresponds to a movieID ranking list of the top 20 most popular movies, annotated horizontally as the rank number and vertically as the user's age group.}
	\centering
	\setlength{\tabcolsep}{1.8mm}
	\begin{tabular}{|l|c|c|c|c|c|c|c|c|c|c|c|c|c|c|c|c|c|c|c|c|}
		\hline
		\diagbox[width=1.2cm,trim=l]{Age}{Rank} &1&2&3&4&5&6&7&8&9&10&11&12&13&14&15&16&17&18&19&20\\
		\hline 1&288&50&181&258&313&121&7&117&172&174&588&333&268&403&302&300&318&298&204&475\\
		\hline 18&50&181&288&56&172&100&258&98&174&127&7&12&1&173&64&313&117&168&210&294\\
		\hline 25&50&1&100&174&98&181&258&127&172&56&173&89&7&79&168&222&151&117&183&318\\
		\hline 35&50&100&127&98&174&258&181&1&286&357&79&318&191&132&515&64&483&56&427&28\\
		\hline 45&100&50&286&258&127&181&300&174&515&9&435&275&15&172&197&204&276&357&199&483\\
		\hline 50&50&286&100&127&313&258&302&269&181&9&300&98&275&237&197&285&194&172&124&318\\
		\hline 56&100&286&127&269&302&285&191&9&313&194&50&199&242&174&300&427&523&275&56&134\\
		\hline
	\end{tabular}
\end{table*}
\begin{figure*}[!t]
	\begin{center}
		\includegraphics[width=5in,height=2in]{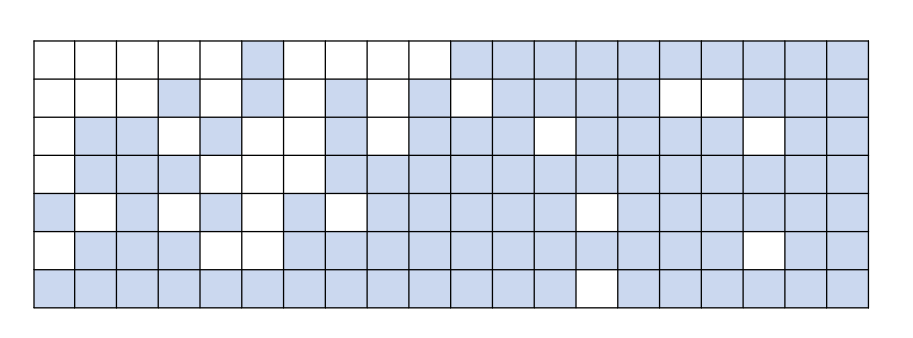}
		\caption{The movie repeat graph corresponding to users of different ages, where the vertical shows different age stages, and the horizontal shows the 20 movies favored by the corresponding users in ascending order from left to right, and the white part shows the globally similar movies.}
	\end{center}
\end{figure*}	

\subsection{Q2:Reasons for using age as user information.}
Some researchers have proved that the use of item types can effectively improve the accuracy of the recommendation system and alleviate the problem of item cold start\cite{b26}. Therefore, this paper makes the following analysis on the use of age factors as user side information:

\subsubsection{Analyze the user side information of the recommendation system}	

\indent\indent In order to alleviate the user cold start problem faced by the recommendation system, the system will use registration information for analysis, and demographic information accounts for a large category of registration information, including user age, gender and so on. In this paper, gender and age are respectively counted, and the proportion of gender and age of users is used for pie chart analysis, and Figure 8 is obtained. In Figure 8, (a) and (b) respectively represent the gender proportion in datasets ml-100k and ml-1m. (c) and (d) respectively represent the proportion of different age stages in the datasets ml-100k and ml-1m, where 1 represents users aged 1-18 years, and so on for other age stages.

As can be seen from the analysis in Figure 8, gender data cannot well reflect the differences between users, while age factors can better distinguish the characteristics of users. As can be seen from the pie chart, most of the data in Movielens dataset are from users aged 25-34 years, and most of them are female, which provides an analysis direction for future research on recommendation system.

\subsubsection{The analysis is carried out from the aspects of user age and recommendation results}

\indent\indent This paper guesses that there are differences in the items preferred by users of different ages and consistency in the items preferred by users of the same age. In order to verify this hypothesis, this paper takes ml-100k dataset as an example to analyze the seven age stages one by one, and find out the top 20 movies most popular among users in different age stages. Table VII lists the top 20 movies with the highest user rating in each stage. In order to better analysis of different age stages of user preference difference of the movie, according to the Table VII created a movie repeated figure, as shown in Figure 7, each row represents the userID for each age group, each column represents the corresponding highest rated movieID, and the white section represents the highest rated repeat movie for each age group.

\subsubsection{The analysis is carried out from the aspects of user age and recommendation results}

\indent\indent This paper guesses that there are differences in the items preferred by users of different ages and consistency in the items preferred by users of the same age. In order to verify this hypothesis, this paper takes ml-100k dataset as an example to analyze the seven age stages one by one, and find out the top 20 movies most popular among users in different age stages. Table VII lists the top 20 movies with the highest user rating in each stage. In order to better analysis of different age stages of user preference difference of the movie, according to the Table VII created a movie repeated figure, as shown in Figure 7, each row represents the userID for each age group, each column represents the corresponding highest rated movieID, and the white section represents the highest rated repeat movie for each age group.
\begin{figure}[H]
	\begin{center}
		\subfigure[ml-100k]{
		\includegraphics[width=1.5in,height=1.25in]{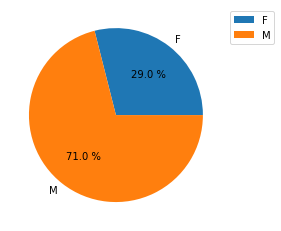}
		}
		\subfigure[ml-1m]{
			\includegraphics[width=1.5in,height=1.25in]{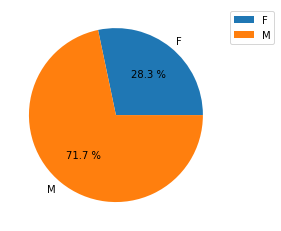}
		}
		\subfigure[ml-100k]{
			\includegraphics[width=1.5in,height=1.25in]{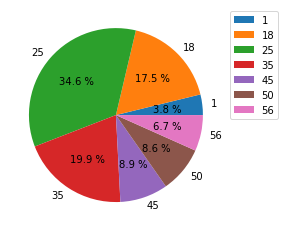}
		}
		\subfigure[ml-1m]{
			\includegraphics[width=1.5in,height=1.25in]{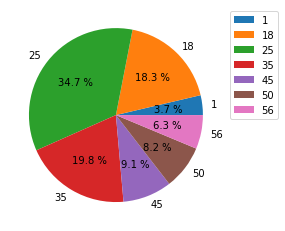}
		}
		\caption{The proportion of gender and age of users. (a) and (c) represent the proportion of gender and age of data set ml-100k. (b) and (d) represents the proportion of different age stages in the dataset ml-1m.}
	\end{center}
\end{figure}
The following two conclusions can be drawn from the above Table VII and Figure 7: (1) as the growth of the age, the differences between different age stages of user preference for the movie is more and more obvious, namely the same preference in gradually reduce the number of movies, such as the first age stage and the second age the same film has nine in user preferences, if only observed the first age stage and the seventh age the same movie has four user preferences, There's only one that corresponds to all the same. (2) There is a certain repetition rate among the user preferences corresponding to different ages. From the total repetition rate, Raiders of the Lost Ark (1981) is favored by users of all ages, and Star Wars (1977) is favored by people aged 1-55.

In order to find out the reason for its repetition, this paper guesses that it is related to the exposure and popularity of the movie. Therefore, this paper starts from the popularity of the movie. For the project, considering its score and viewing times, this paper defines its popularity as:
\begin{equation}
	{\rm{popularity}}\left( i \right){\rm{ = }}\left( {\frac{1}{2}grade\left( i \right) + \frac{1}{2}unums\left( i \right)} \right) \times 100\%
\end{equation}
where $grade\left( i \right) = \frac{{The{\rm{ }}rating{\rm{ }}of{\rm{ }}i}}{{The{\rm{ }}total{\rm{ }}score}}$, $grade\left( i \right)$ represents the ratio of the total score of the item i to the total score of the whole items. $unums\left( i \right) = \frac{{views{\rm{ }}number\left( i \right)}}{{The{\rm{ }}total{\rm{ }}records}}$, $unums\left( i \right)$ represents the ratio of the viewing times of an item to the total record, use weights to combine the two aspects. After calculating the popularity of the movies, we sorted them and found the top 10 movies that were most popular among users, as shown in Table IX.
\vspace{10pt}
\begin{table}
	\caption{A list of the top 10 most popular movies, including the movie's rank number, MovieID, title, and corresponding popularity.}
	\centering
	\setlength{\tabcolsep}{2pt}
	\begin{tabular}{c|c|c|c}
		\hline ranking&MovieID&Title&Popularity\\
		\hline 1&50&Star Wars (1977)&3.2608\%\\
		\hline 2&100&Fargo (1996)&2.7090\%\\
		\hline 3&258&Contact (1997)&2.5589\%\\
		\hline 4&174&Raiders of the Lost Ark (1981)&2.2920\%\\
		\hline 5&127&Godfather, The (1972)&2.2701\%\\
		\hline 6&286&English Patient, The (1996)&2.2573\%\\
		\hline 7&1&Toy Story (1995)&2.2496\%\\
		\hline 8&98&Silence of the Lambs, The (1991)&2.1469\%\\
		\hline 9&288&Scream (1996)&2.1110\%\\
		\hline 10&56&Pulp Fiction (1994)&2.0533\%\\
		\hline
	\end{tabular}
\end{table}

Table IX shows that the most popular movie among users is Star Wars (1977), with a popularity of 3.2608\%. By comparing Table IX with Table VII, it is obvious that users of different ages are all affected by the popularity of movies. Excluding the top ten movies with the highest influence, the film repetition chart can be obtained again, as shown in Figure 9(a). In order to analyze the details, this paper counted the most preferred movies of users aged 1-34 and above 50, and made the movie repeat Figure 9(b), which can also prove that there are great differences in the degree of preference of users at different ages, and only three movies are the same. This also shows that the user's preference degree is deeply affected by the popularity of the item, so the future recommendation system can improve the influence of popularity on the recommendation system, so as to achieve the purpose of personalized recommendation.
\begin{figure}[htp]
	\begin{center}
		\subfigure[Seven age stages]{
			\includegraphics[width=2.5in,height=1in]{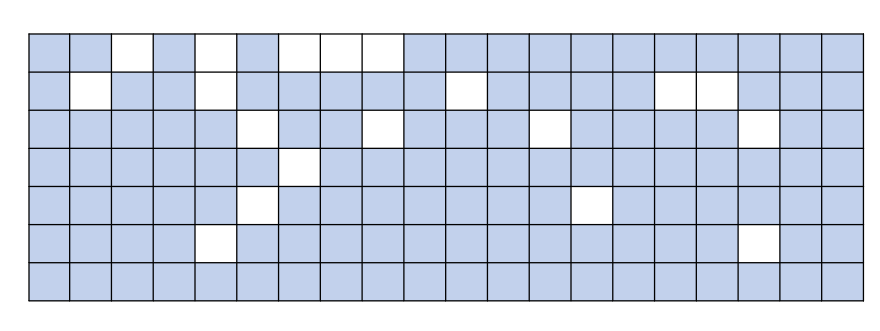}
		}
		\subfigure[Ages 1-34 and over 50]{
		\includegraphics[width=2.5in,height=0.5in]{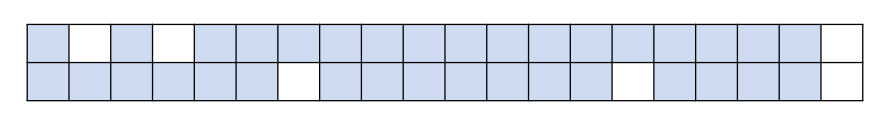}
	}
		\caption{Movie repeats affected by popularity are excluded. (a) represents the film repetition chart of seven stages, and (b) is the film repetition chart of users aged from 1 to 34 and above 50.}
	\end{center}
\end{figure}

\section{Conclusion}
This paper proposes a matrix decomposition model based on user-item related attributes, which projects the age attributes of each user and the type attributes of each movie into the same potential factor space as the user and item, enriching the item representation and user representation in MF. At the same time, it is combined with SVD++ model to integrate not only the related attributes of users and items, but also the implicit representation of users and items. According to the experimental analysis of UISVD++ model on ml-100k and ml-1m, the results prove that the means of integrating potential user age and film type attributes into SVD++ framework is effective. In addition, the performance of our model on two datasets is better than that of the given deep learning hybrid recommendation model, and also better than other algorithms.

Studies show that UISVD++ helps to improve data sparsity and cold priming problems with adjustable. However, since only one attribute of user and item is included respectively, the effect of UISVD++ is not as good as the model with deep learning when the proportion of training data is 50\%. As a future work, we plan to add richer item and user attributes to further improve the recommendation performance of UISVD++, and promote UISVD++ to the recommendation system in various situations. In addition, integrating the temporal dynamic factors into UISVD++ makes the model not only has the attributes of users and items, but also can change the recommendation with time, so as to improve the practicability of the recommendation.\\

\noindent \textbf{Data Availability}\\
https://grouplens.org/datasets/movielens/100k/  \\
https://grouplens.org/datasets/movielens/1m/ \\




\begin{thebibliography}{00}
\bibitem{b1} H. Shan and A. Banerjee, “Generalized probabilistic matrix factorizations for collaborative filtering,” in Proc. 2010 IEEE Int. Conf. Data Mining, 2010, pp. 1025–1030.
\bibitem{b2} Z. Cui, X. Xu, F. Xue, X. Cai, Y. Cao, W. Zhang, and J. Chen, “Personalized recommendation system based on collaborativefiltering for IoTscenarios,” IEEE Trans. Services Comput., vol. 13, no. 4, pp. 685–695,Jul. 2020.
\bibitem{b3} Y. Koren, R. Bell, and C. Volinsky, “Matrix factorization techniques for recommender systems,” Computer, vol. 42, no. 8, pp. 30–37, Aug. 2009.
\bibitem{b4}Y. Shi, M. Larson, and A. Hanjalic, “Collaborativefiltering beyond the user-item matrix: A survey of the state of the art and future challenges,” ACM Comput. Surv., vol. 47, no. 1, pp. 1–45, Jul. 2014.
\bibitem{b5}  J. Bobadilla, F. Ortega, A. Hernando, and A. Gutiérrez, “Recommender systems survey,” Knowl.-Based Syst., vol. 46, pp. 109–132, Jul. 2013.
\bibitem{b6} X. Su and T. M. Khoshgoftaar, “A survey of collaborativefiltering techniques,” Adv. Artif. Intell., vol. 2009, pp. 1–19, Oct. 2009.
\bibitem{b7} I. Fernández-Tob´ıas, I. Cantador, P . Tomeo, V . W. Anelli, and T. Di Noia, “Addressing the user cold start with cross-domain collaborative filtering: exploiting item metadata in matrix factorization,” User Modeling and User-Adapted Interaction, vol. 29, no. 2, pp. 443–486, 2019.
\bibitem{b8} R. P. Adams, G. E. Dahl, and I. Murray, “Incorporating side information in probabilistic matrix factorization with Gaussian process,” in Proc. 26th Conf. UAI, Catalina Island, CA, USA, Jul. 2010, pp. 1–9.
\bibitem{b9} T. Zhao, J. McAuley, and I. King, “Leveraging social connections to improve personalized ranking for collaborativefiltering,” in Proc. 23rd ACM Int. Conf. Conf. Inf. Knowl. Manage., Shanghai, China, Nov. 2014, pp. 261–270.
\bibitem{b10} Z. Cheng, X. Chang, L. Zhu, R. C. Kanjirathinkal, and M. Kankanhalli, “MMALFM: Explainable recommendation by leveraging reviews and images,” ACM Trans. Inf. Syst., vol. 37, no. 2, pp. 1–28, Mar. 2019.
\bibitem{b11} Z. Cheng, Y. Ding, X. He, L. Zhu, X. Song, and M. Kankanhalli, “A3NCF: An adaptive aspect attention model for rating prediction,” in Proc. 27th Int. Joint Conf. Artif. Intell., 2018, p. 3748–3754.
\bibitem{b12} J. Nguyen and M. Zhu, “Content-boosted matrix factorization techniques for recommender systems,” Stat. Anal. Data Mining, vol. 6, no. 4, pp. 286–301, Aug. 2013.
\bibitem{b13} F. M. Harper and J. A. Konstan, “The MovieLens datasets: History and context,” ACM Trans. Interactive Intelligent Systems, vol. 5, no. 4, pp.1–19, 2015.
\bibitem{b14} G. Guo, J. Zhang, and N. Yorke-Smith, “A novel recommendation model regularized with user trust and item ratings,” IEEE Trans. Knowl. Data Eng., vol. 28, no. 7, pp. 1607–1620, Jul. 2016.
\bibitem{b15} M. Jamali and M. Ester, “A matrix factorization technique with trust propagation for recommendation in social networks,” in Proc. 4th ACM Conf. Recommender Syst. (RecSys), New York, NY, USA, Sep. 2010, pp. 135–142.
\bibitem{b16} S. Zhang, L. Yao, A. Sun, and Y. Tay, “Deep learning based recommender system: A survey and new perspectives,” ACM Comput. Surveys, vol. 52, no. 1, pp. 1–38, Feb. 2019.
\bibitem{b17} M. F. Dacrema, P. Cremonesi, and D. Jannach, “Are we really making much progress? A worrying analysis of recent neural recommendation approaches,” in Proc. 13th ACM Conf. Recommender Syst., New York, NY, USA, Sep. 2019, p. 101–109.
\bibitem{b18} S. Rendle, L. Zhang, and Y. Koren, “On the difficulty of evaluating baselines: A study on recommender systems,” 2019, arXiv:1905.01395. [Online]. Available: http://arxiv.org/abs/1905.01395
\bibitem{b19} Z. Sun, Q. Guo, J. Yang, H. Fang, G. Guo, J. Zhang, and R. Burke, “Research commentary on recommendations with side information: A survey and research directions,” Electron. Commerce Res. Appl., vol. 37, Sep. 2019, Art. no. 100879.
\bibitem{b20}S. Funk, “Netflix update: Try this at home,” 2006.
\bibitem{b21} Y. Koren, “Factorization meets the neighborhood: A multifaceted collaborative filtering model,” in Proc. 14th ACM SIGKDD Int. Conf. Knowl. Discovery Data Mining (KDD), Las Vegas, NV, USA, 2008, pp. 426–434.
\bibitem{b22} R. Salakhutdinov and A. Mnih, “Probabilistic matrix factorization,” in
Proc. 20th Int. Conf. Neural Inf. Process. Syst. Red Hook, NY, USA: Curran Associates, 2007, pp. 1257–1264.
\bibitem{b23} Rui Duan, Cuiqing Jiang, Hemant K. Jain, “Combining review-based collaborative filtering and matrix factorization: A solution to rating's sparsity problem,”Decision Support Systems, vol.156, pp.113748, February 2022.
\bibitem{b24} Hai Liu, Chao Zheng, Duantengchuan Li, et.al. “EDMF: Efficient Deep Matrix Factorization With Review Feature Learning for Industrial Recommender System,” IEEE TRANSACTIONS ON INDUSTRIAL INFORMA TICS, vol. 18, pp.4361-4371, July 2022.
\bibitem{b25} Yaxing Lai, Wanbo Yu, “Recommendation system based on user preference and prediction rating,”ISCIPT, vol.978, pp:185-189, July 2021.
\bibitem{b26} Haiyang Zhang, Ivan Ganchev, Nikolov, et al.“FeatureMF: An Item Feature Enriched Matrix Factorization Model for Item Recommendation,”IEEE Access, vol. 9, pp.65266-65276, April 2021.
\bibitem{b27} H Wang, Z Hong, M Hong, “Research on product recommendation based on matrix factorization models fusing user reviews,”Applied Soft Computing, vol.123, pp.1568-4946, July 2022.
\bibitem{b28} Zhiguo Zhu, Mengru Yan, et.al, “Rating prediction of recommended item based on review deep learning and rating probability matrix factorization,” Electronic Commerce Research and Applications, vol.54, pp.1567-4223, Aug 2022.
\bibitem{b29} Suvash Sedhain, Aditya Krishna Menon, et al. “AutoRec: Autoencoders Meet Collaborative Filtering,”Association Computing Machinery, vol.978, no. 1, pp.4503-3473, May, 2015.
\bibitem{b30} Hang Zheng, Xing Xing, Qiuyang Han, et al. “UIAE: Collaborative Filtering for User and Item based on Auto-Encoder,”ICNISC, vol.978, no.1, pp.6654-0232, March, 2021.
\bibitem{b31} Y. Zhang, C. Zhao, et al. “Integrating Stacked Sparse Auto-Encoder Into Matrix Factorization for Rating Prediction,” IEEE Access, vol.9, pp.17641-17648, Sep 2021.
\bibitem{b32} Z. Zamanzadeh Darban, et al. “GHRS: Graph-based Hybrid Recommendation System with Application to Movie Recommendation,”Expert Systems with Applications, vol.200, pp.116850-116864, August 2022.
\bibitem{b33} T. Chai and R. R. Draxler, “Root mean square error (RMSE) or mean absolute error (MAE)?–Arguments against avoiding RMSE in the literature,” Geoscientific Model Develop., vol. 7, no. 3, pp. 1247–1250, Jun. 2014.
\bibitem{b34} H. Zhang, N. S. Nikolov, and I. Ganchev, “Exploiting user feedbacks in matrix factorization for recommender systems,” in Model and Data Engineering. Cham, Switzerland: Springer, Sep. 2017, pp. 235–247.
\bibitem{b35} Steffen Rendle, “Factorization machines,” In 2010 IEEE International Conference on Data Mining. IEEE, 995–1000.
\bibitem{b36}  L. Yu, W. Shuai, and M. S. Khan, “A novel deep hybrid recommender system based on auto-encoder with neural collaborative filtering,” Big Data Mining Anal., vol. 1, no. 3, pp. 211–221, 2018.

\end{thebibliography}
\end{document}